\documentclass{article}
\usepackage{ijcai23}
\usepackage{xcolor}

\usepackage{times}
\usepackage{soul}
\usepackage{url}
\usepackage[hidelinks]{hyperref}
\usepackage[utf8]{inputenc}
\usepackage[small]{caption}
\usepackage{graphicx}
\usepackage{amsmath}
\usepackage{amsthm}
\usepackage{booktabs}
\usepackage{algorithm}
\usepackage{algorithmic}
\usepackage[switch]{lineno}
\usepackage[
    backend=biber,
    style=numeric,
    sorting=none
  ]{biblatex}
\title{HoneyGAN Pots: A Deep Learning Approach for Generating Honeypots}

\author{
Ryan Gabrys
\and
Daniel Silva\And
Mark Bilinski
\affiliations
Naval Information Warfare Center Pacific\\
\emails
\{ryan.c.gabrys, daniel.silva61, mark.bilinski\}.civ@us.navy.mil
}

\begin{document}

\maketitle

\begin{abstract}
    This paper investigates the feasibility and effectiveness of employing Generative Adversarial Networks (GANs) for the generation of decoy configurations in the field of cyber defense. The utilization of honeypots has been extensively studied in the past; however, selecting appropriate decoy configurations for a given cyber scenario (and subsequently retrieving/generating them) remain open challenges. Existing approaches often rely on maintaining lists of configurations or storing collections of pre-configured images, lacking adaptability and efficiency. In this pioneering study, we present a novel approach that leverages GANs' learning capabilities to tackle these challenges. To the best of our knowledge, no prior attempts have been made to utilize GANs specifically for generating decoy configurations. Our research aims to address this gap and provide cyber defenders with a powerful tool to bolster their network defenses. 
    \end{abstract}

\section{Introduction}

The field of cybersecurity constantly faces the challenge of defending networks and systems against malicious attacks. One effective approach to deceive adversaries and gather intelligence about their tactics is through the use of decoy systems, which are commonly known as honeypots \cite{stoll,Rowe}. By strategically deploying honeypots at different stages of the cyber kill chain, organizations can gain valuable insights into the attacker's methods, motives, and vulnerabilities \cite{FW20}. Honeypots can provide early warning signs, capture attack tools or malware samples, and gather valuable threat intelligence that can enhance overall security.

Honeypots are typically categorized as being either high or low-interaction depending on their level of sophistication. Low-interaction honeypots typically target an attacker at the earlier phases of the cyber kill chain, such as  reconnaissance, whereas the high-interaction honeypots aim to disrupt potential attackers at later phases, such as delivery and lateral movement \cite{ZT21}.  Although the methods described in this work can be applied to generate either low-interaction or high-interaction honeypots, our primary focus in this work is on the design and generation of realistic-looking low-interaction honeypots that can be used in the context of a cyber defense strategy to detect, deter and/or delay potential attackers.


\subsection{Our Approach}
Our proposed method utilizes an adaptable infrastructure that only requires two essential pieces of information: the services available on each open port and the operating system details. We note that such infrastructures   have existed for some time \cite{P04},\cite{NCSS15} but one of the challenges that still exists is determining which decoy configurations to choose from, as well as how the configurations should be generated. Some naive approaches involve maintaining a list of possible device configurations or in some cases even storing collections of pre-configured images. 

Our approach harnesses the capabilities of GANs to learn the distribution of network device configurations using real data. This approach offers remarkable flexibility, as cyber defenders no longer need to maintain collections of potential configurations. Instead, our GAN-powered system dynamically generates realistic-looking decoy configurations based on specified requirements. Particularly in scenarios where a large number of diverse and authentic-looking decoys are desired, this methodology has the potential to offer enormous benefit.

The main objective of this paper is to explore the feasibility and effectiveness of using GANs for generating decoy configurations in cyber defense, which, to the best of the author's knowledge, has not been attempted before. This work represents a first effort where future works will incorporate additional information about the network environment that can be used to better inform the design of honeypots. By leveraging the learning capabilities of GANs, we aim to provide cyber defenders with a powerful tool to enhance their network defenses.

\subsection{Contributions}

Our contributions are the following:

\begin{enumerate}
\item  Using a simple data model, we show that a GAN can generate high-quality replicas of actual network device configurations using the concepts of precision and recall from \cite{SBLBG15}.
\item For the setup where a cyber defender wishes to generate certain types of decoys, we design two conditional GANs that generate network device configurations based upon operating system or service type.

\item We demonstrate that the resulting decoys created from our generative models are robust to current honeypot detection systems.
\end{enumerate}

\subsection{Outline}

Section~\ref{sec:background} reviews the concepts of Generative Adversarial Networks (GANs) as well as the use of honeypots for network defense. In Section~\ref{sec:ganar}, we present the models for our GAN architecture and Section~\ref{sec:datamodel} reviews our simple data model. In Section~\ref{sec:eval}, we measure both the diversity and accuracy of the GANs that generated and trained using real-world data. Finally, Section~\ref{sec:conclude} concludes the paper.

\section{Background and Related Work}\label{sec:background}

\subsection{Generative Adversarial Networks}
Generative Adversarial Networks (GANs) have gained significant attention due to their 
ability to generate synthetic data simulating realistic media such as images, text, audio and videos. GANs were introduced in 2014 by Ian Goodfellow \cite{GPAMXWFOCB14} and have since sparked a revolution in the field of generative modeling. 
Unlike traditional generative models which are typically trained by maximizing a log likelihood, GANs possess a unique architectural setup consisting of two neural networks: the generator and the discriminator. 

The generator network takes random noise as input and generates synthetic samples, such as images, text, or even audio. The discriminator network, on the other hand, receives both real and generated samples and tries to distinguish between them. The two networks are trained together in a competitive setting, constantly improving and challenging each other's performance. More specifically, the discriminator, denoted as $D$, and generator, denoted as $G$, play the following two-player min-max game with value function $V(G,D):$
\begin{align}
\min_{G} \max_{D} V(D,G) = &E_{x \sim p_r(x)} \left[ \log D(x) \right] \nonumber \\
&+ E_{z \sim p_g(z)} \left[ \log\left( 1-D(G(z)) \right) \right],
\end{align}
where $p_g$ represents the generator's distribution and $p_z$ represents a prior on input noise variables. The generator aims to generate samples that are indistinguishable from real data, while the discriminator strives to correctly classify between real and fake samples. As training progresses, the generator learns to produce increasingly realistic samples by receiving feedback from the discriminator. The discriminator, in turn, becomes more adept at distinguishing between real and generated data.

The applications of GANs span a wide range of domains. In computer vision, GANs have been used for image synthesis, style transfer, image-to-image translation, and super-resolution \cite{GPAMXWFOCB14}, \cite{IZZE17}. They have also been employed in generating realistic deepfake videos and enhancing image generation in areas like fashion, art, and design \cite{GD18}. In natural language processing, GANs have been utilized for text generation, language translation, and dialogue systems. GANs have even found applications in healthcare, where they have been used for generating synthetic medical images, augmenting data for training medical models, and drug discovery \cite{FADKAGG18}. 

Nowadays, GANs are broadly studied and applied through academic and industrial research in different domains beyond 
media (e.g., natural language processing, medicine, electronics, networking, and cybersecurity) \cite{PBBBW17}, \cite{PLKPKH23}. After a GAN has been trained, its generator can produce 
as many synthetic examples as necessary, providing an efficient mechanism for solving the problem of lack of labelled data 
sets and potential privacy restrictions.

\subsection{Cyber Defensive Deception}

Honeypots are decoy systems or resources intentionally designed to attract and deceive malicious actors, allowing cybersecurity professionals to study their behavior, gather intelligence, and enhance their defensive strategies. Dating back to at least as far as the early 1980s, researchers have considered strategies involving staging imaginary computer environments (honeypots) to lure attackers and reveal their objectives \cite{stoll}.

The Honeynet Project, initiated by Lance Spitzner and a group of security professionals, was one of the pioneering efforts in developing and deploying honeypots. The project aimed to capture and analyze the tactics, techniques, and tools used by hackers.  These can be used to provide valuable insights into honeypot deployment strategies and the analysis of captured attack data \cite{S03}.

Later works considered various methods used by attackers to detect and evade honeypots. In order to address many of the challenges, in \cite{PCZ19} the authors emphasized the importance of continuous monitoring and adaptation to stay ahead of sophisticated adversaries. The work in \cite{GPYF07} focuses on using honeypots as a means to gather intelligence on botnets, which are networks of compromised computers controlled by malicious actors. It demonstrates how honeypots can be used to detect, monitor, and analyze botnet activities, providing valuable information for improving network defenses and disrupting cybercriminal operations.

In \cite{SI15}, the use of honeypots to analyze and understand Advanced Persistent Threat (APT) campaigns is considered. APTs are stealthy and prolonged cyber attacks typically orchestrated by nation-state actors or advanced criminal organizations. One of the many advantages of employing honeypots in these environments is that they can provide insights into attacker tactics, helping organizations improve their defense mechanisms against such threats \cite{BLM14}.

\section{Experimental Setup and GAN Architecture}\label{sec:ganar}

We instantiated our GANs with the following properties:
\begin{itemize}
\item \textbf{Batch Size}: This refers to the number of machine configurations that are used for training at each iteration of the optimizer. We used a batch size of $64$.
\item \textbf{Number of steps}: Each step involves a single generator iteration along with $3$ discriminator training iterations. Our unconditional GAN was trained for $11844$ steps whereas our conditional O/S GAN and conditional DT GAN were trained for $5922$ and $23688$ steps, respectively.
\item \textbf{Gradient Penalty Coefficient}: This represents the loss term that keeps the L2 norm of the discriminator gradients close to $1$. We set this parameter to $10$.
\item \textbf{Adam optimizer hyper parameters}: 
\begin{itemize}
\item \textbf{Learning Rate}: Controls how quickly parameters in the model are updated.
\item \textbf{Coefficient $\beta_1$, $\beta_2$}:  Manages the decay rates of the moving average of the gradient and the squared gradient. 
\end{itemize}
For our setup, we set the learning rate to be $0.0002$ and the values of $\beta_1, \beta_2$ to be $0.5, 0.9$ respectively.
\end{itemize}

\begin{figure}[h]
\includegraphics[width=8cm]{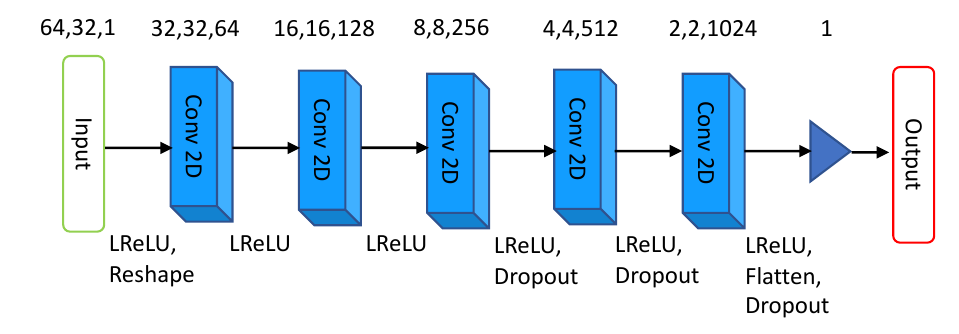}
\vspace{-2.0ex}
\caption{Discriminator Model}
\label{fig:Disc}
\end{figure}

Our architecture consists of two neural networks where, as is illustrated in Figure~\ref{fig:Disc}, the discriminator network is using fractionally strided convolutions and the generator network is performing a deconvolution procedure. The discriminator model has a normal convolution layer followed by four convolution layers using a stride of $2$ and a kernel of size $5$ to downsample the input. The final layer of the discriminator is a dense layer that provides a single output. 

\begin{figure}[h]
\includegraphics[width=8cm]{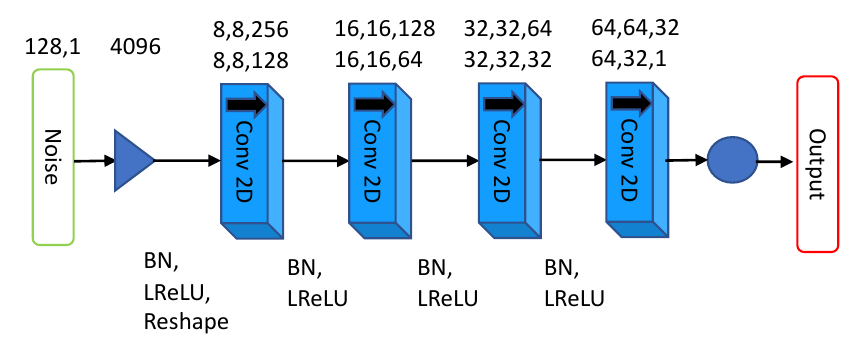}
\caption{Generator Model}
\label{fig:Gen}
\end{figure}

The generator model consists of four blocks where each block performs upsampling followed by a convolution layer that  has a kernel of size $3$ with a stride of size $1$. After these four blocks, the data is passed through a final activation layer. Both models were trained according using the Wassterstein GAN with Gradient Penalty method \cite{WGAN}.

\section{Dataset and Model}\label{sec:datamodel}
\begin{figure}[h]
\includegraphics[width=8cm]{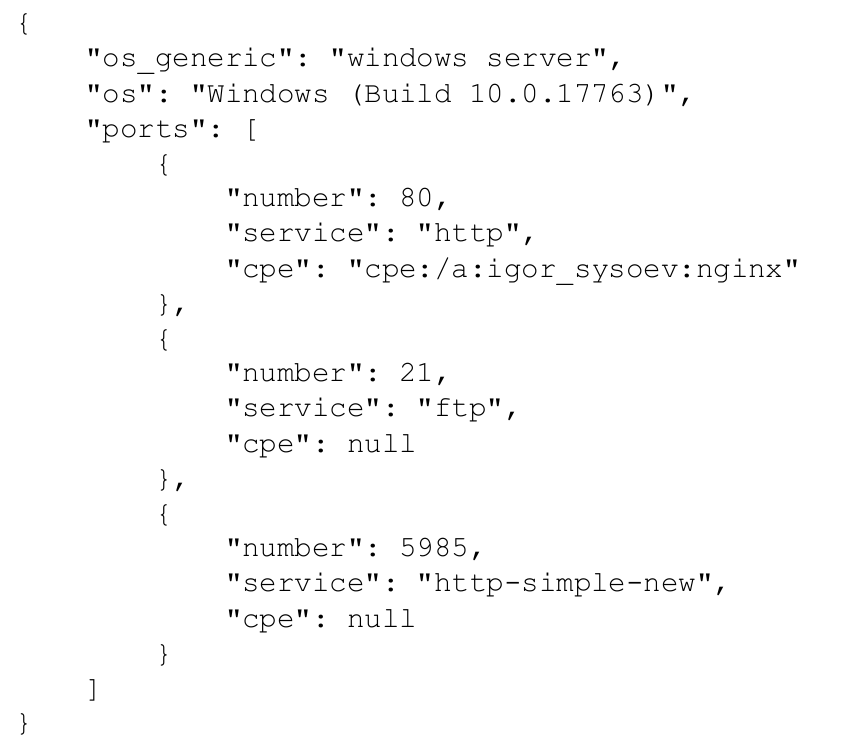}
\caption{JSON representation of Windows Server device}
\label{fig:json}
\end{figure}

For the purpose of evaluating our proposed approach, we trained the GAN described in the previous section using a set of device configurations that were obtained using the Shodan search engine. Using Shodan, we were able to extract the configurations of $378973$ internet-connected devices. Each configuration consisted of the set of open ports, the service running on each of those ports, along with any Common Platform Enumeration (CPE) data that is associated with the particular service. For the purposes of illustrating this data, Figure~\ref{fig:json} shows an example JSON document representing the device configuration for a Windows Server running three services.

\begin{figure}[h]
\includegraphics[width=8cm]{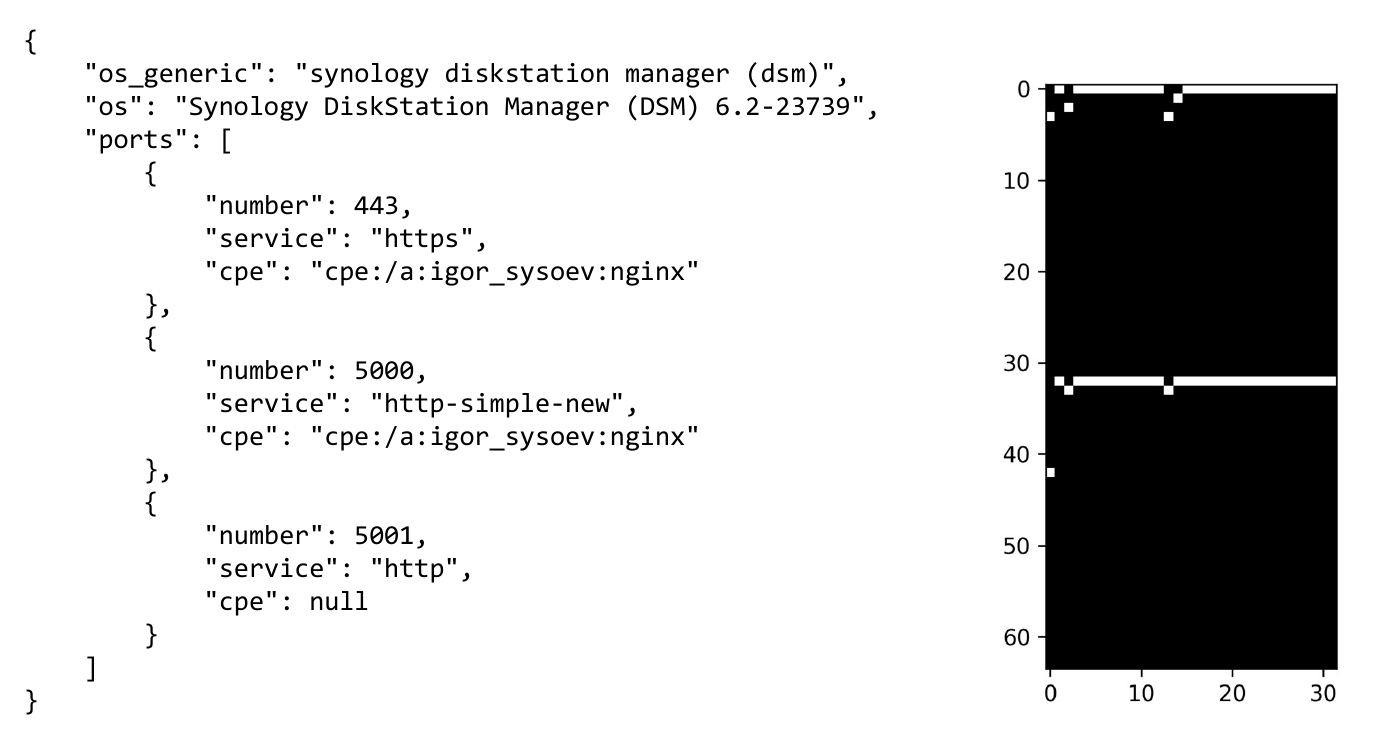}
\caption{Sample Data Model for Machine Configuration}
\label{fig:datamodel}
\end{figure}

As can be seen in Figures~\ref{fig:Disc} and \ref{fig:datamodel}, each device configuration is represented as a $64x32$ object. The first two columns (which contains $64$ rows) encodes the information associated with the operating system and build associated with the particular device. The remaining $30$ columns contain information pertaining to the services. Under this model, and as is being illustrated in Figure~\ref{fig:datamodel}, each column will contain exactly two ones.\footnote{The ones in each column of Figure~\ref{fig:datamodel} are colored white.} 

\begin{figure*}[h]
\centering
\includegraphics[scale=0.55]{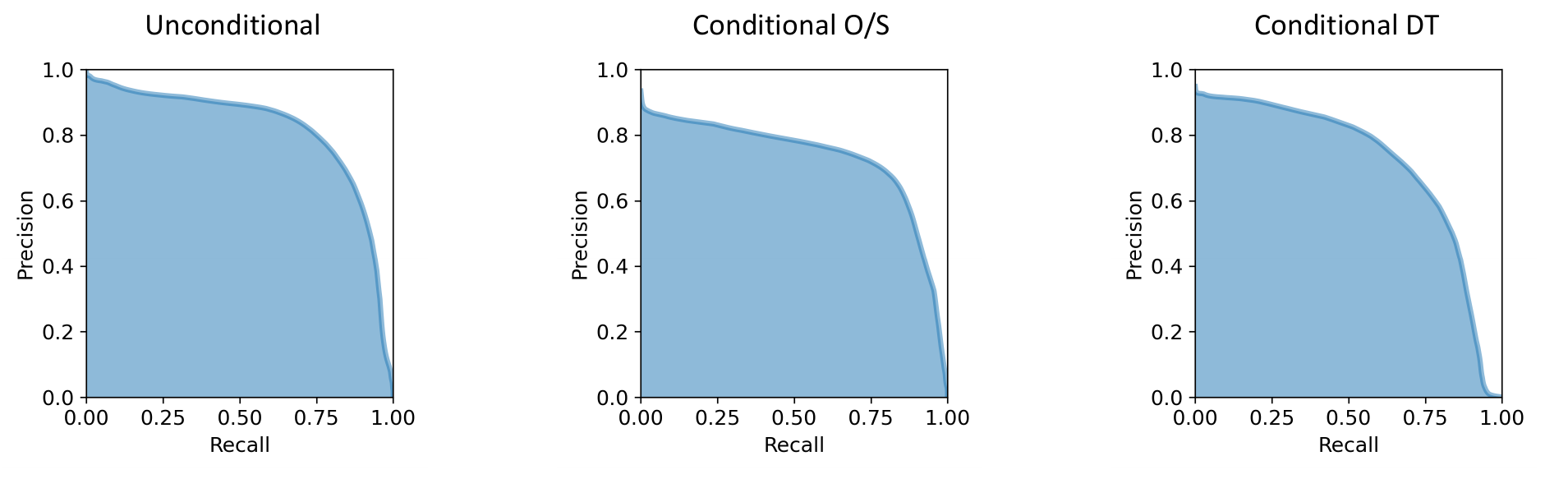}
\center{\caption{PRD Plots for Achievable $(\alpha, \beta)$ Pairs Across Different GANs}\label{fig:data}}
\end{figure*}

Each of the next $30$ columns (after the first two) represent a particular port. The assignment of columns to ports as well as the methodology in selecting these $30$ ports, is described in more detail in Appendix~\ref{app:psdm}.  Each of these $30$ columns contains exactly two ones where the one in the first $32$ rows indicates the service that is running and the second one present in rows $33-64$ indicates the CPE associated with the service. 

\section{Evaluation}\label{sec:eval}

In order to evaluate the performance of our GAN, we leveraged the notion of precision and recall originally introduced in \cite{SBLBG15}. Conceptually, \textit{precision} measures how accurately the generated samples are to the true reference distribution whereas \textit{recall} attempts to capture how well the true sample distribution is ``covered'' by the generated distribution. More precisely, suppose that $P_R$ represents a reference distribution and $P_G$ represents the generated (or learned) distribution. Then for $\alpha, \beta \in [0,1]$, we say that the probability distribution $P_G$ has precision $\alpha$ and recall $\beta$ with respect to $P_R$ if there exists distributions $\mu, \nu_{P_R}, \nu_{P_G}$ if 
\begin{align*}
    P_R &= \beta \mu + (1-\beta) \nu_{P_R}\\
    P_G &= \alpha \mu + (1-\alpha) \nu_{P_G}.
\end{align*}
As can be seen from the expressions above, the parameter $1-\beta$ is the recall loss whereas $1-\alpha$ is the loss due to precision. The set of attainable pairs of precision and recall of a distribution $P_{G}$ with respect to $P_{R}$ consists of all $(\alpha, \beta)$ pairs that are achievable.


Figure~\ref{fig:data} is illustrating the PRD plots for the three types of GANs that were developed, which we refer to as our unconditional GAN, conditional O/S GAN, and conditional Device Type (DT) GAN. The left plot of Figure~\ref{fig:data} displays the $(\alpha,\beta)$ pairs for our unconditional (unlabeled) GAN. The unconditional GAN was trained for $11844$ steps after which time the performance of the GAN did not improve. In general, and as can be seen above, this GAN typically exhibited the highest achievable $(\alpha,\beta)$ pairs. As a concrete example, if we fix the level of recall to be $0.75$, we see that a precision of $\alpha > 0.8$ is possible. 

Despite its high levels of precision and recall, our unconditional GAN has the drawback that a human cyber defender has no control over the types of samples that it outputs. In order to overcome this potential issue, we investigated the performance of two additional conditional GANs:
\begin{itemize}
    \item Conditional O/S GAN: This model was trained using labels that reflected the operating system. 
    \item Conditional Device Type (DT) GAN: This model was trained using labels reflecting the function or type of the underlying device. Details of how these device type categories were assigned to devices can be found in Appendix~\ref{app:cdtgan}.
\end{itemize}
For the conditional O/S GAN, each device was assigned exactly one of the following labels reflecting its O/S: Mikrotik Routeros, Windows Server, Windows, Synology Diskstation Manager, Sonicwall Sonico, Linux, Ubuntu, Debian, Synology Router Manager, or QTS. The exact distribution of these labels to our data set is included in Appendix~\ref{app:psdm}.

For the conditional DT GAN, each device was assigned one or more of the following labels: file sharing, remote access, webserver, mailserver, database, dns, vpn, router, management. The procedure that was followed in assigning devices to labels under the DT GAN architecture is described in more detail in Appendix~\ref{app:cdtgan}. The performance of our conditional GANs were evaluated similarly by computing the respective PRD plots. Recall that the conditional DT GAN was trained for $23688$ steps whereas the conditional O/S GAN was  trained for $5922$ steps.\footnote{Similar to the unconditional GAN, training the GANs for more steps did not improve the performance.}

As can be seen from Figure~\ref{fig:data}, although the conditional GANs afford a cyber defender greater control over the output configuration (or samples), the penalty for such control are lower levels of achievable $(\alpha, \beta)$ pairs. For example, the conditional O/S GAN never achieves a precision above $0.80$ when the recall is above $0.55$ and similarly the conditional DT GAN does not achieve a precision above $0.75$ when the recall is held at $0.75$. Interestingly, the conditional DT had higher values of precision when the recall was allowed to be smaller, but the conditional O/S exhibited larger levels of precision for larger recall values.

In order to better understand the performance of our GAN architecture when provided a smaller number of generated samples, we considered the quality of the unconditional GAN as a function of the number of outputs. In particular, Table~1 is displaying the number of distinct port/service/os combinations that are generated from our GAN as a function of the number of samples requested. For these outputs, we sampled 
with replacement from the set of real configurations that were retrieved from Shodan as well as a set of configurations that were generated by our unconditional GAN. For example, the first row of the table is showing that among  $500$ randomly selected real configurations, there were $406$ whose port/service/os information was unique.  Then, after our generator produced an equivalent $500$ samples, there were $295$ unique samples and $247$ matched a real configuration. As the generator could generate a given sample multiple times, some of the matches may be repeats of the same configuration, which is why in many cases there are a larger number of matches than unique configurations and is likely a result of over-fitting.  As expected, as the number samples increases, the diversity of outputs from our GAN also decreases. Similar trends were observed with respect to the other two GANs and this data is included in Appendix~\ref{app:fconditionals}.  
\begin{table}[h!]
\label{tab:fgen}
  \begin{center}
    \begin{tabular}{l|c|c|c} 
      \textbf{Samples} & \textbf{Real, Unique} & \textbf{Gen, Unique} & \textbf{Gen, Match}  \\
      \hline
      500 & 406 & 295 & 247 \\
      1000 & 720 & 465 & 536 \\
      1500 & 1006 & 623 & 750 \\
      2000 & 1319 & 702 & 999 \\
      2500 & 1602 & 837 & 1280 \\
      3000 & 1881 & 916 & 1470 \\
      3500 & 2109 & 969 & 1812 \\
      4000 & 2351 & 1075 & 1951 \\
      4500 & 2603 & 1105 & 2337 \\
      5000 & 2844 & 1209 & 2514
    \end{tabular}
        \caption{Generator Output as Function of Sample Size for Unconditional GAN}
  \end{center}
\end{table}

In addition to considering how the samples generated from each of our three GANs compared to the real set, we also leveraged HoneyD to generate actual low-interaction decoys using port/service/OS data from our unconditional GAN. Using these decoys, we then evaluated the quality of the resulting decoys using the Checkpot \cite{checkpot} honeypot checker utility, which was developed for the purposes of helping security researchers check that their honeypots are properly set up in such a manner as to attract ``high-quality traffic.'' Towards this end, the Checkpot utility works by taking the network address of a honeypot as input and then outputting a number, known as a Karma value, that indicates the quality of the honeypot specified as input. We created $5000$ low-interaction honeypots with HoneyD using $5000$ randomly generated configurations from our unconditional GAN. The  average Karma value for the resulting collection of honeypots was $491.584$. As a baseline, we also evaluated the Karma value of another publicly available low-interaction honeypot, which according to Checkpot, had a Karma value of only 60 \cite{masscanned}. In comparison, we generated HoneyD decoys using $5000$ randomly selected real device configurations and found the average Karma value to be $504.91$, indicating that the quality of honeypots is nearly the same as if we had stored hundreds of thousands of actual machine configurations.

\section{Conclusion}\label{sec:conclude}

In this paper, we have considered the feasibility of applying deep learning techniques to the problem of generating realistic-looking decoy configurations. We have shown that a GAN can generate high-quality replicas of actual network device configurations using precision and recall metrics, and that this result can be achieved through the use of a relatively simple data model. Additionally, two conditional GANs were designed to generate network device configurations based on either operating system or service type. The resulting decoys produced by these generative models were then demonstrated to be resilient against current honeypot detection systems.

Despite the progress made in our research, it is important to acknowledge the limitations or potential drawbacks of the approach considered. One of the potential shortcomings of our approach is the relatively sparse data representation that was chosen. Future works will consider the use of embeddings in the hopes of yielding a more efficient representation. Another aspect of our design was that duplicate samples were used to train both the generator and discriminator models, which likely led to over-fitting, and potentially negatively impacted the diversity of our resulting model.

Moving forward, there are several promising avenues for further research in this field. Building upon the findings and insights gained from our study, future investigations could focus on the following areas: (a) Development of an efficient embedding for representing machine configurations along with the configurations of other machines in the same network environment, (b) Design of a decoy generation system whose output depends on other factors in the environment such as the configurations of other machines in the subnet or data about the presence or behaviors of suspected attackers, (c) Development of generative models for producing alternative types of honeytokens to aid in deception and network defense.

\appendix

\section{Representation of Port/Service Information in Data Model}\label{app:psdm}
As mentioned in Section~\ref{sec:datamodel}, our two-dimensional data representation uniquely associates columns $2,3,\ldots,31$ to represent the following ports: 
\begin{align*}
&21, 22, 23, 25, 53, 80, 110, 123, 135, 137, 139, 161, 443, 445, \\
&1433, 1701, 1723, 2000, 3306, 3389, 4433, 5000, 5001, 5985, \\
&8080, 8081, 8291, 8443, 8728, 9100.
\end{align*}
These ports represent the most frequently occurring open ports that were found across the set of $378973$ Shodan device configurations that were used to train the models described in Section~\ref{sec:ganar}. On average, this data representation captured over $85\%$ of the services running on each device.  Table~2 is displaying the number of configurations in our dataset for each of the $9$ possible types of operating systems.

\begin{table}[h!]
\label{table:oslabels}
  \begin{center}
    \begin{tabular}{l|c} 
      \textbf{O/S Label} & \textbf{Count} \\
      \hline
      MikroTik Routeros & 119478  \\
      Windows Server & 104584  \\
      Windows & 50391 \\
      DiskStation Manager & 49609 \\
      SonicOS & 25489 \\
      Linux & 9781 \\
      Ubuntu  & 8960 \\
      Synology Router Manager & 3616\\
      Debian  & 5391 \\
      QTS & 1674
    \end{tabular}
        \caption{Labeling Operating Systems}
  \end{center}

\end{table}

\section{Device Types in Conditional DT GAN}\label{app:cdtgan}

In order to allow a user to specify the type(s) of device generated by our conditional DT GAN, we assigned one or more labels to each of the device configurations used during the training process described in Section~\ref{sec:ganar}. The logic behind this assignment is being illustrated in Table~2 below. For each device, we labeled the device to be of the type listed in the first column if the module name that was listed in the data field provided by Shodan contained at least one of the substrings contained in the second column. For instance, if the module string returned by Shodan for a particular device contains the substring `http,' then according to our procedure the device would be a webserver. We note that under this procedure, a device may be assigned multiple device type labels so that it is possible that a device is both a webserver and a file sharing server if, for instance, the device is running both an http service and an ftp service. The number of each device type is also represented in the third column.

\begin{table}[h!]
\label{tab:dtlabel}
  \begin{center}
    \begin{tabular}{l|c|c} 
      \textbf{Device Type} & \textbf{Substrings} & \textbf{Count} \\
      \hline
      webserver & \{http\} & 296999  \\
      file sharing & \{smb, ftp\} & 110127 \\
      mailserver & \{imap, pop3, smtp\} & 21550 \\
      database & \{sql\} & 27975 \\
      management & \{ldap, snmp, ntp, kerberos\} & 35161 \\
      dns & \{ `dns' \} & 40936 \\
      remote access & \{telnet, ssh, rdp\} & 136116 \\
      vpn & \{pptp, l2tp, openvpn\} & 42824 \\
      router & \{router\} & 70949
    \end{tabular}
        \caption{Labeling Device Types}
  \end{center}

\end{table}

\section{Conditional Generator Outputs for Finite Sample Sizes}\label{app:fconditionals}

\begin{figure}[h]
\centering
\includegraphics[width=8cm]{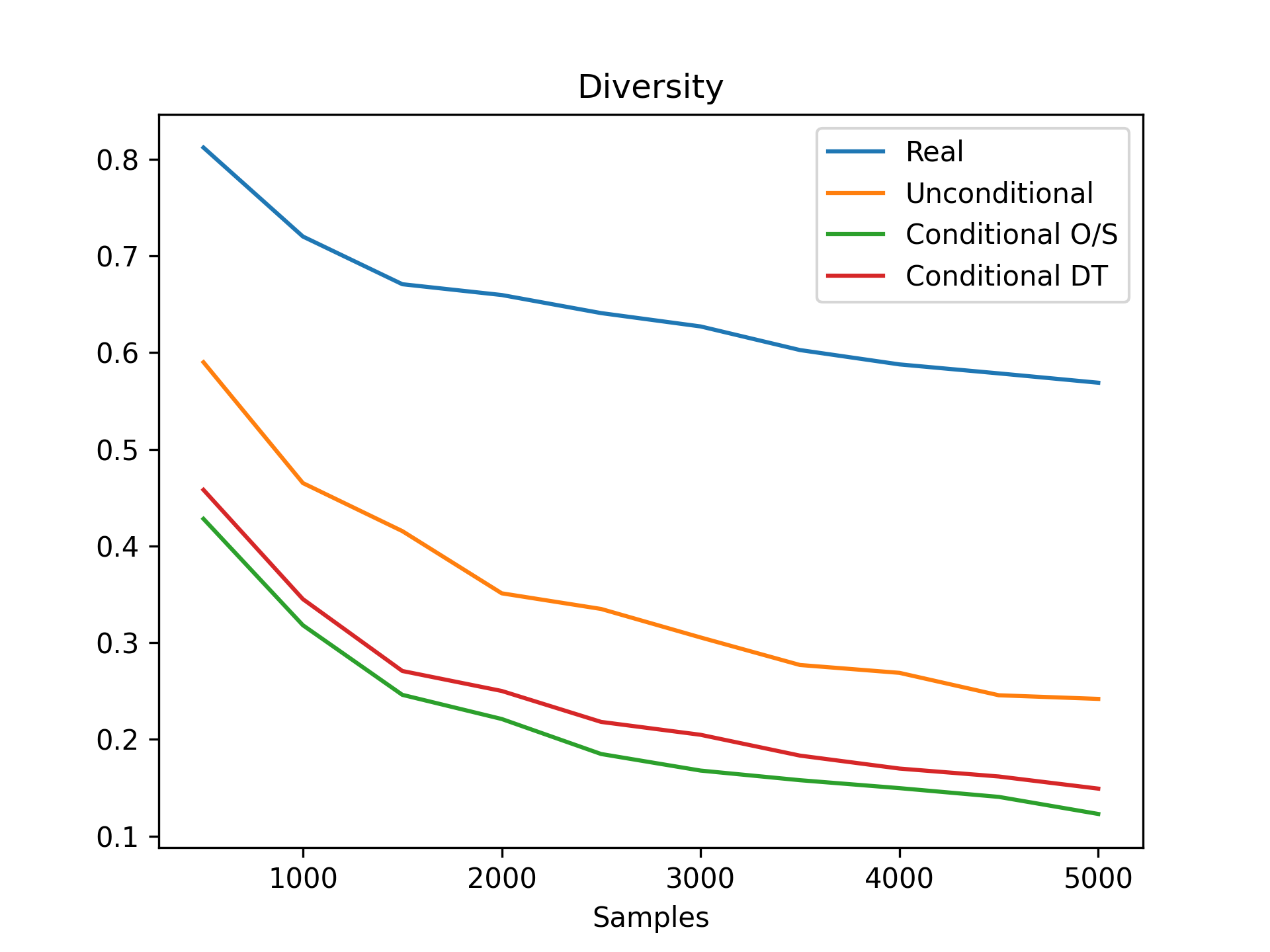}
\caption{Fraction of Generated Samples that are Unique}
\label{fig:divGANs}
\end{figure}

\begin{figure}[h]
\centering
\includegraphics[width=8cm]{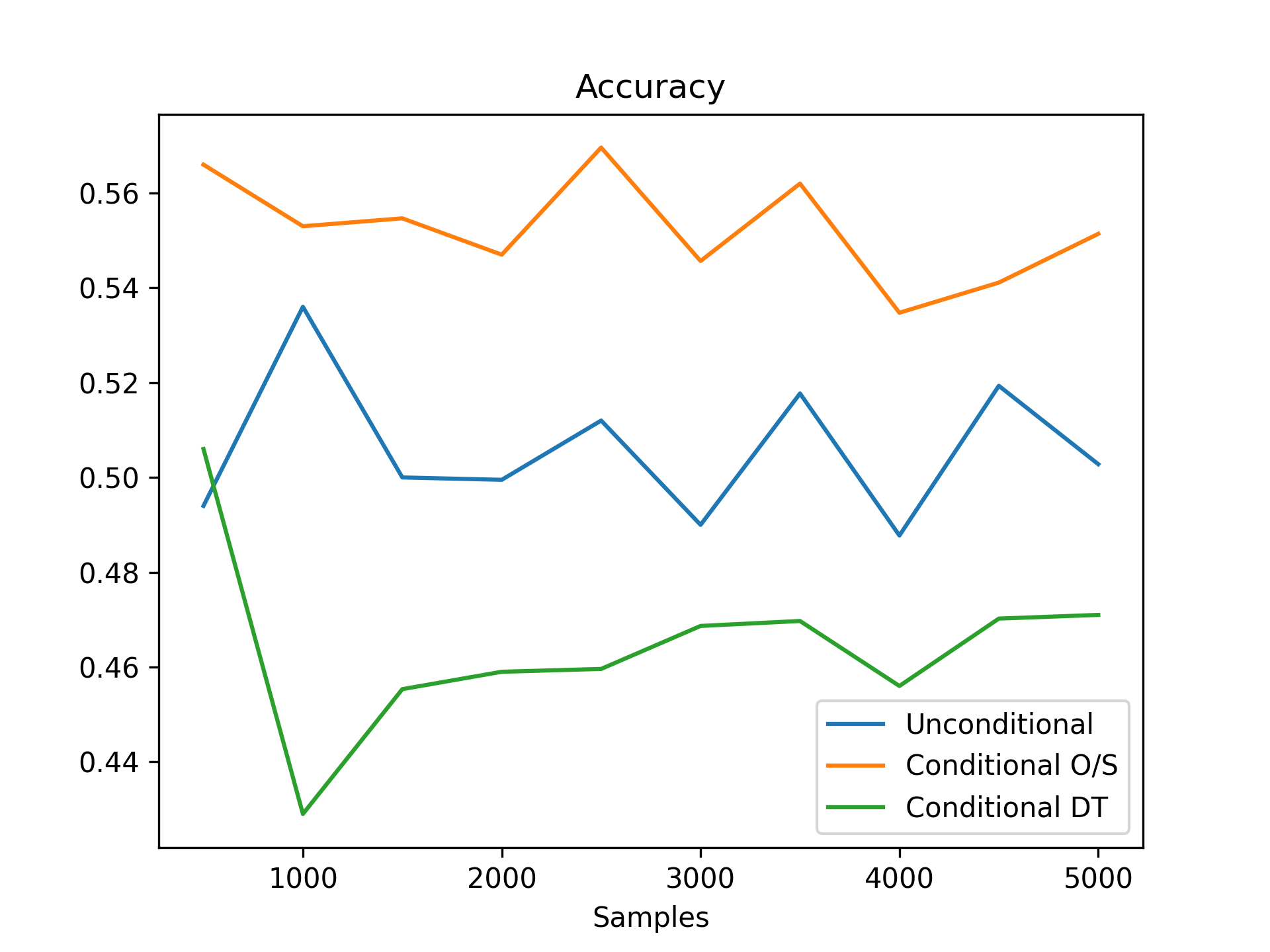}
\caption{Fraction of Generated Samples that Match Real Samples}
\label{fig:accGANs}
\end{figure}

Figure~\ref{fig:divGANs} is showing the fraction of unique samples that were generated by each of our GANs as a function of the number of total samples requested. Note that the data from Table~1 is represented as the top two lines of our graph. For instance, Table~1 is indicating that given $1000$ samples (x-axis) generated from our unconditional GAN the fraction of unique samples is $465/1000$ which is displayed as the second line from the top in Figure~\ref{fig:divGANs}. Information pertaining to the accuracy of each of the three GANs is being displayed in an analogous manner in Figure~\ref{fig:accGANs}.

\printbibliography

@book{ stoll,
  author = "C. Stoll",
  title = "The Cuckoo’s egg: Tracking a spy through the maze of computer espionage",
  publisher = "Doubleday",
  year = "1989"
}

@inproceedings{ Rowe,
  author = "N.C. Rowe",
  title = "A model of deception during cyber-attacks on information systems",
  booktitle = "IEEE First Symposium on Multi-Agent Security and Survivability",
  pages = "21--30",
  publisher = "IEEE",
  address = "Drexel, PA",
  month = "August",
  year = "2004"
}

@book{ FW20,
  author = "K. Ferguson-Walter",
  title = "An Empirical Assessment of the Effectiveness of Deception for Cyber Defense",
  publisher = "University of Massachusetts Amherst",
  year = "2020"
}

@inproceedings{NCSS15,
  author = "B. Nagpal et al.",
  title = "Catch: Comparison and analysis of tools covering honeypots",
  booktitle = "International Conference on Advances in Computer Engineering and Applications",
  publisher = "IEEE",
  address = "Ghaziabad, India",
  pages = "783--786",
  month = "July",
  year = "2015"
}

@inproceedings{P04,
  author = "N. Provos",
  title = "A virtual honeypot framework",
  booktitle = "Proceedings of the 13th Conference on USENIX Security Symposium",
  publisher = "IEEE",
  address = "San Diego, CA",
  month = "August",
  year = "2004"
}

@inproceedings{SBLBG15,
  author = "M.S.M. Sajjadi et al.",
  title = "Assessing
generative models via precision and recall",
  booktitle = "Advances in Neural Information
Processing Systems",
  address = "Montreal, Canada",
  pages = "5228–-5237",
  month = "December",
  year = "2015"
}

@article{ZT21,
  title={Three decades of deception techniques in active cyber defense-retrospect and outlook},
  author={Zhang, Li and Thing, Vrizlynn LL},
  journal={Computers \& Security},
  volume={106},
  pages={102288},
  year={2021},
  publisher={Elsevier}
}

@article{S03,
  author = "L. Spitzner",
  title = "The Honeynet Project: trapping the hackers",
  journal = "Security \& Privacy",
  publisher = "IEEE",
  volume = "1",
  number = "2",
  month = "April",
  pages = "15--23",
  year = "2003"
}

@article{ PCZ19,
  author = "J. Pawlick, E. Colbert, and Q. Zhu",
  title = "A Game-theoretic Taxonomy and Survey of Defensive Deception for Cybersecurity and Privacy",
  journal = "ACM Computing Surveys",
  volume = "52",
  number = "4",
  month = "August",
  pages = "1--28",
  year = "2019"
}

@article{FADKAGG18,
  author = "M. Frid-Adar et al.",
  title = "GAN-based synthetic medical image augmentation for increased CNN performance in liver lesion classification",
  journal = "Neurocomputing",
  volume = "321",
  number = "10",
  pages = "321--331",
  month = "December",
  year = "2018"
}

@inproceedings{IZZE17,
  author = "P. Isola et al.",
  title = "Image-to-image translation with conditional adversarial networks",
  booktitle = "Conference on computer vision and pattern recognition",
  pages = "1125-–1134",
  publisher = "IEEE",
  address = "Honolulu, HI",
  month = "July",
  year = "2017"
}

@inproceedings {SI15,
author = "Z. Saud and M.H. Islam",
title = "Towards proactive detection of advanced persistent threat (APT) attacks using honeypots",
booktitle = "8th International Conference on Security of Information and Networks",
address = "New York, NY",
publisher = "ACM",
month = "September",
year="2015"
}

@inproceedings {BLM14,
author = "V. Bukac, V. Lorenc, and V. Matyáš",
title = "Red Queen’s Race: APT Win-Win Game",
booktitle = "Cambridge International Workshop on Security Protocols",
address = "Cambridge, United Kingdom",
publisher = "Springer",
pages = "55–-61",
month = "March",
year="2014"
}

@inproceedings {GPYF07,
author = "G. Gu et al.",
title = "BotHunter: Detecting Malware Infection Through {IDS-Driven} Dialog Correlation",
booktitle = "16th USENIX Security Symposium (USENIX Security 07)",
address = "Boston, MA",
publisher = "USENIX Association",
month = "August",
year="2007"
}

@inproceedings{GD18,
  author = "D. Güera and E.J. Delp",
  title = "Deepfake Video Detection Using Recurrent Neural Networks",
  booktitle = "International Conference on Advanced Video and Signal Based Surveillance (AVSS)",
  publisher = "IEEE",
  address = "Auckland, New Zealand",
  month="November",
  year = "2018"
}

@inproceedings{GPAMXWFOCB14,
  author = "I.J. Goodfellow et al.",
  title = "Generative Adversarial Nets",
  booktitle = "Advances in Neural Information Processing Systems",
  volume = "27",
  publisher = "Curran Associates, Inc.",
  year = "2014"
}

@article{PBBBW17,
  author = "O. Press et al.",
  title = "Language Generation with Recurrent Generative Adversarial Networks without Pretraining",
  journal = "arXiv 1706.01399",
  year = "2017"
}

@article{PLKPKH23,
  author = "C. Park et al.",
  title = "An Enhanced AI-Based Network Intrusion Detection System Using Generative Adversarial Networks",
  journal = "Internet of Things Journal",
  publisher = "IEEE",
  volume = "10",
  number = "3",
  pages = "2330--2345",
  month = "February",
  year = "2023"
}

@misc{checkpot,
  author = {{The Honeynet Project}},
  publisher = "Google Summer of Code",
  title = {{checkpot}},
  howpublished = {\url{https://checkpot.readthedocs.io/en/master/index.html}},
  year = "2018",
  note = ""
}

@misc{masscanned,
  author = {{Ivre}},
  publisher = "GitHub",
  title = {{Masscanned}},
  howpublished = {\url{https://github.com/ivre/masscanned}},
  year = "2018",
  note = ""
}

@misc{WGAN,
  author = "A.K. Nain",
  title = "WGAN-GP",
  howpublished = {\url{https://keras.io/examples/generative/wgan_gp/}},
  year = "2020",
  note = ""
}
\end{document}